\documentclass[a4paper,11pt]{article}
\usepackage{aaskaiid}
\usepackage{amsmath}

\usepackage{siunitx}
\usepackage{aas_macros}
\usepackage{float}

\usepackage{graphicx}
\usepackage{longtable}

\usepackage{orcidlink}

\usepackage{xeCJK}

\usepackage{xcolor}

\newcommand{\revipz}[1]{}

\title{Probing the solar corona and the solar wind using angular
broadening observations with the SKA}
\ShortTitle{Angular broadening with SKA}

\author[1]{Peijin Zhang (张沛锦) \orcidlink{0000-0001-6855-5799}}
\ShortName{P Zhang} 
\affiliation[1]{Center for Solar-Terrestrial Research, New Jersey Institute of Technology, Newark, NJ 07102, USA}
\emailAdd{peijin.zhang@njit.edu}

\author[2]{John Morgan\orcidlink{0000-0001-9224-5483}}
\affiliation[2]{CSIRO, Space and Astronomy, P.O. Box 1130, Bentley, WA 6102}

\author[3]{Divya Oberoi\orcidlink{(0000-0002-4768-9058)}}
\affiliation[3]{National Centre for Radio Astrophysics, Tata Institute of Fundamental Research, Pune 411007, India.}

\author[4]{Du Toit Strauss\orcidlink{0000-0002-0205-0808}}
\affiliation[4]{Center for Space Research, North-West University, Potchefstroom, 2531, South Africa}

\author[5]{Yingjie Luo\orcidlink{0000-0002-5431-545X}}
\author[5]{Eduard Kontar\orcidlink{0000-0002-8078-0902}}
\affiliation[5]{University of Glasgow, Glasgow, United Kingdom, UK}

\author[6]{Zesen Huang\orcidlink{0000-0001-9570-5975}}
\affiliation[6]{Department of Earth, Planetary, and Space Sciences, University of California, Los Angeles }

\author[7]{Keshav Aggarwal\orcidlink{0000-0002-7004-8670}}

\author[8]{Anshu Kumari\orcidlink{0000-0001-5742-9033}}
\affiliation[8]{Udaipur Solar Observatory, Physical Research Laboratory, Dewali, Badi Road, Udaipur - 313001, Rajasthan, India}

\author[7]{Abhirup Datta\orcidlink{0000-0002-5333-1095}}

\affiliation[7]{Department of Astronomy, Astrophysics and Space Engineering (DAASE), Indian Institute of Technology Indore, Indore, Madhya Pradesh 453552, India}

\author[9]{Diana E. Morosan\orcidlink{0000-0002-8416-1375}}
\affiliation[9]{Department of Physics and Astronomy, University of Turku, 20014, Turku, Finland}

\author[10]{Gert J. J. Botha\orcidlink{0000-0002-5915-697X}}
\affiliation[10]{School of Engineering, Physics and Mathematics, Northumbria University, Newcastle upon Tyne, NE1 8ST, UK}

\abstract{Angular broadening observations of compact radio sources provide a powerful method for probing the solar corona and solar wind. These observations enable the study of various properties such as the phase structure-function, amplitude of turbulence, middle-scale density variations, solar wind heating rates, and dissipation scales. When a compact radio source is observed through the solar coronal plasma, it can result in several observable effects: 1. The size of the radio source increases due to scattering by the turbulent plasmas in the solar corona or solar wind, known as angular or scatter broadening. 2. The flux density of the source drops due to scattering and absorption. 3. Further, the observed angular broadening exhibits anisotropy, which reflects the anisotropic nature of the coronal and solar wind turbulence responsible for the scattering. 4. The position angle (PA) of the observed anisotropies, measured from north through east, can help infer the orientation of the magnetic field within the solar corona. These phenomena provide insights into the physical processes governing the solar wind and its interaction with electromagnetic waves, offering constraints on coronal turbulence and magnetic field configurations. Currently, we have a limited number of studies on the angular broadening, and mostly on very bright sources such as Tau-A. 
However, it is expected that the SKA's unprecedented resolution and sensitivity will provide many more radio source candidates, thereby offering greater insights into our understanding of the solar corona, solar wind, and the heliosphere.}

\begin{document}
\maketitle

\section{Introduction}

\begin{figure}[h]
    \centering
    \includegraphics[width=1.06\linewidth]{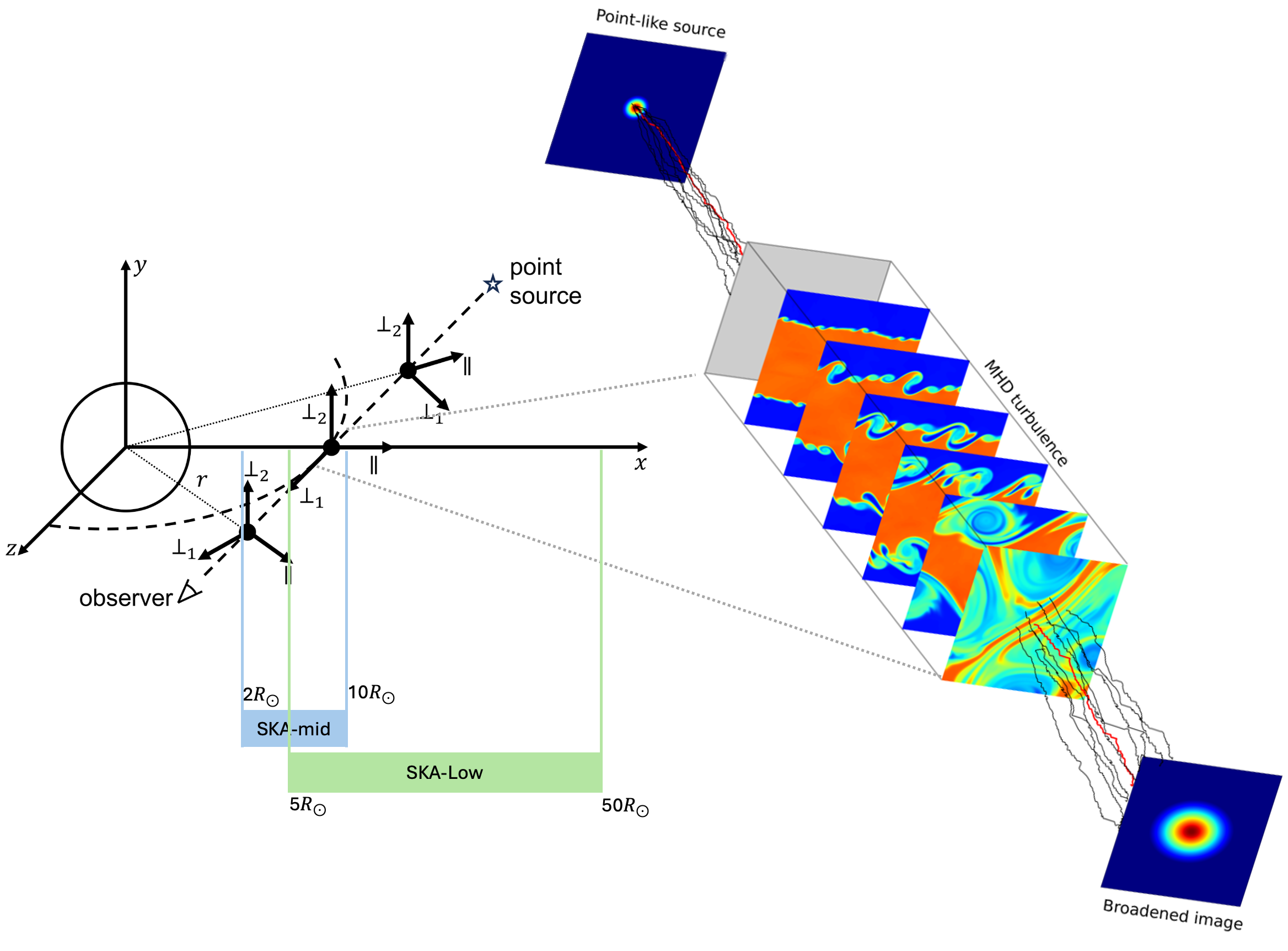}
    \caption{Concept demonstration of angular broadening observation for the turbulence in the plasma of the corona and solar wind. Rays from a compact background source (right) propagate along the line of sight  pass the corona then reach the observer (left figure adapted from \citealt{kontar2023anis}). Density fluctuations from MHD turbulence on the line of sight will convert the point-like source into a broadened, elongated image, as shown on the right side of the figure.
    SKA-Mid covers 2-10 $R_\odot$ corona, and SKA-Low covers 5-50 $R_\odot$ corona.
    }
    \label{fig:concept}
\end{figure}

The higher corona and the inner heliosphere (2-50 $R_\odot$) plasma parameters are difficult to measure either remotely or \textit{in situ}. Spacecraft traverse only limited heliocentric distances and latitudes and can only provide measurements at given locations, while optical observation diagnostics often suffer from the low emissivity of this region. Angular broadening observations of compact background radio sources (conceptual geometry shown in Figure  \ref{fig:concept}) provide a complementary and powerful approach: phase fluctuations imposed by coronal/solar-wind turbulence blur and anisotropically stretch the apparent images of distant sources, allowing the density--fluctuation spectrum, anisotropy, and its evolution with heliocentric distance to be inferred directly along the line of sight.
Recent analyses have shown that the observed source shapes and their position-angle alignment are well explained by \emph{anisotropic} scattering in magnetized turbulence, with the elongation tracing the projected coronal field direction \citep[e.g.][]{kontar2019anis,kontar2023anis}. 
The measurements of the broadened size, axial ratio, and orientation---tracked as a function of relative location of the Sun---become quantitative probes of the fluctuation amplitude and profile and the dissipation (inner) scale of turbulence.

\begin{figure}
    \centering
    \includegraphics[width=1.05\linewidth]{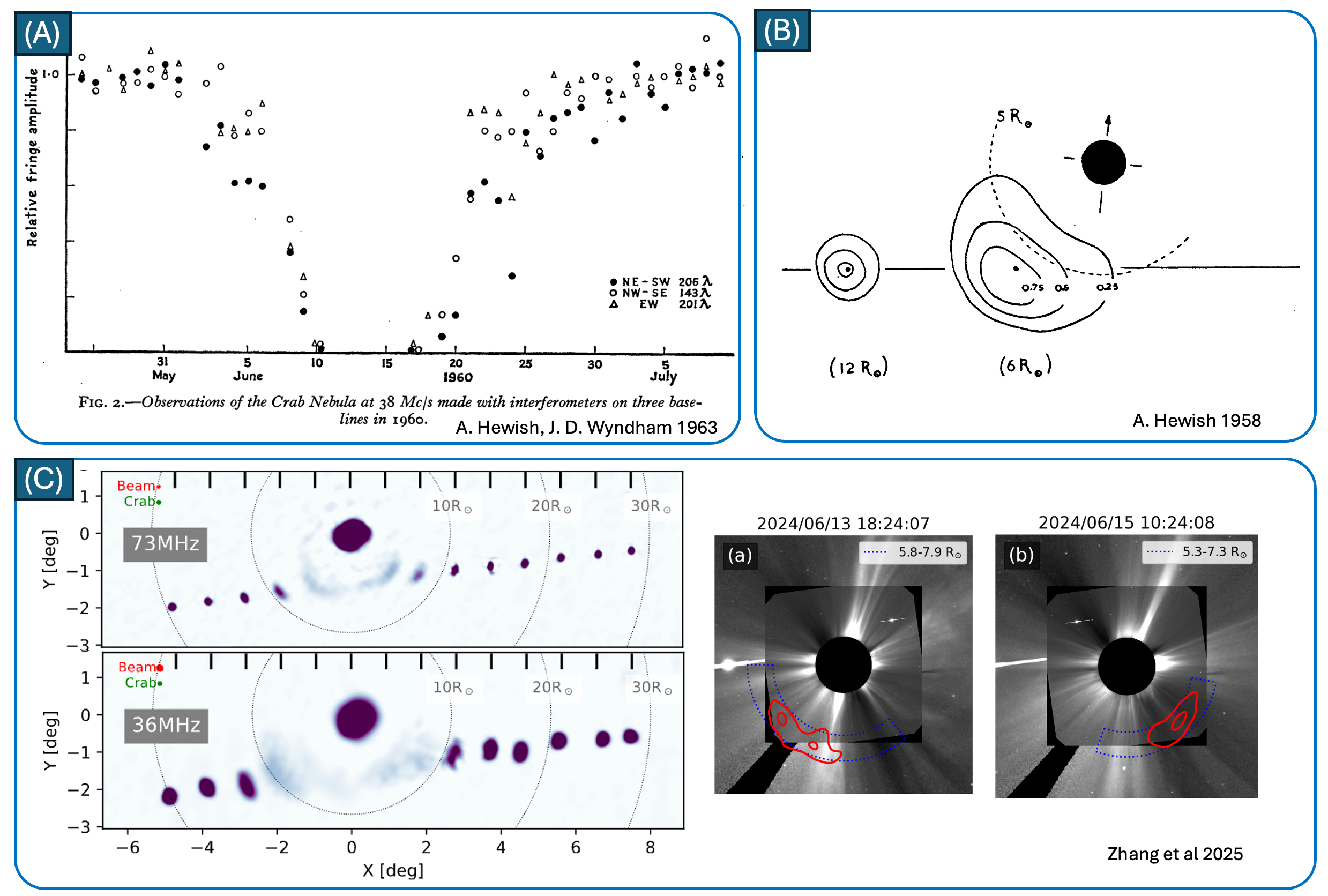}
    \caption{Previous studies in angular-broadening studies of background sources near the Sun. (A) Early interferometric detections of the Crab Nebula scattering at 38 MHz \citep{hewish1963solar} showing baseline-dependent fringe suppression near solar conjunction. (B) Concept sketch of anisotropic coronal scattering whose elongation follows the coronal magnetic field \citep{hewish1958scattering}. (C) Recent low-frequency imaging example (36 and 73 MHz): the Crab's apparent size and orientation evolve as it approaches the Sun (top), with contemporaneous coronagraph context at comparable elongations  \citep{2025zhangTauA}. Together, these panels trace the historical to current evidence for strong, anisotropic coronal/solar-wind scattering.}
    \label{fig:milestones}
\end{figure}

The feasibility and value of angular-broadening measurements have been demonstrated over six decades of observations. 
Pioneering interferometric experiments already showed fringe suppression of the Crab Nebula near solar conjunction (panel A Fig \ref{fig:milestones}; \citealt{1963HewishMNRASsolar}) and motivated an anisotropic-scattering picture in the magnetized corona (panel B Fig \ref{fig:milestones}; \citealt{hewish1958scattering}). At low radio frequencies, a large fraction of campaigns have targeted Tau A (Crab Nebula) because of its exceptional brightness and its annual passage close to the Sun, which provides a compact, stable beacon for scattering experiments. Meter--decameter observations have mapped the growth of the Crab Nebula's apparent size and its strong anisotropy versus solar elongation and position angle, and have tracked the associated flux depression during ingress/egress. Recent imaging further resolves the evolving, elongated Crab Nebula near conjunction and relates it to contemporaneous coronal structure (panel C; \citealt{2025zhangTauA}). Taken together, these studies show that (i) the observables are robust with high-fidelity, direction-dependent calibration; (ii) the anisotropy position angle closely follows the large-scale coronal magnetic-field orientation; and (iii) multi-epoch, multi-source campaigns can recover the radial evolution of the turbulence spectrum and inner (dissipation) scale.
T

The Square Kilometre Array (SKA) now offers a significant step forward.
The dense core improves the visibility sampling of scatter-broadened structures, while the long baselines are more sensitive to the small angular structures that may be intrinsic. Together, these features enable precise characterization of the coronal scattering profile, anisotropy, and turbulence spectrum.
The increased sensitivity enables detection of weaker and more distant compact sources, thereby improving the constraints on turbulence and scattering.
Measurements with the AAVS2 prototype station and end-to-end simulations indicate array sensitivities of order $A_{\rm eff}/T_{\rm sys}\sim10^3  \mathrm{m^2 K^{-1}}$ under cold-sky conditions in the \SIrange{70}{160}{MHz} range \citep{macario2022characterization,2021SKA_AAVS_PASAsokoloski}. 
SKA-Mid, operating at higher frequencies (\SI{350}{MHz} to \SI{15}{GHz}), extends the capability for coronal and solar-wind angular-broadening studies to higher angular resolution and probes smaller inner turbulence scales. Its increased sensitivity and longer baselines will enable angular broadening and scattering measurements of compact sources in regions farther from the Sun and at different coronal heights, complementing the capabilities of SKA and providing a broader frequency baseline for characterizing solar and heliospheric turbulence.
This capability greatly benefits angular-broadening experiments: measurements are limited by calibration and dynamic range rather than thermal noise in the vicinity of the Sun, allowing deeper direction-dependent calibration and solar peeling. \revipz{Here, “solar peeling” refers to modeling and subtraction of the bright solar emission in the visibility domain and its sidelobes so that weaker background sources near the Sun can be imaged reliably.} The dense $uv$ sampling and long baselines resolve sub-arcminute, anisotropically broadened shapes. These advances allow routine, high-cadence, multi-source measurements of coronal/solar-wind turbulence with SKA.

In this paper, we assess the prospects for SKA angular-broadening studies. 
Section \ref{sec:1} presents the theoretical foundation and corresponding observables of angular broadening observations.
Section \ref{sec:2} quantifies background-source availability along the Sun's annual track using low-frequency sky surveys. Section \ref{sec:3} translates SKA station/array sensitivities into image-domain RMS limits and dynamic-range requirements in solar proximity, accounting for the elevated $T_{\rm sky}$ and solar contamination. Section \ref{sec:4} the observation plans and strategies according to Sections \ref{sec:2} and \ref{sec:3}.

\section{Angular broadening observables}
\label{sec:1}

For a compact, far-field radio source, the signals received by an 
interferometer (two antennas) on a projected baseline $\mathbf{s}$ traverse the scattering 
medium along lines of sight separated by exactly the same distance $s=|\mathbf{s}|$.  
Density irregularities introduce phase fluctuations 
$\phi(\mathbf{r})$, and the relevant statistical quantity is the 
{phase structure function}
$
D_\phi(\mathbf{s}) \equiv 
\left\langle \left[\phi(\mathbf{r}+\mathbf{s})-\phi(\mathbf{r})\right]^2 \right\rangle .
$
The mutual coherence of the wavefield measured by the interferometer is
$\Gamma(\mathbf{s}) = {V(\mathbf{s})}/{V(\mathbf{0})}
    = \exp(-D_\phi(\mathbf{s})/2)$
so a single complex visibility at baseline $s$ is a direct probe of 
$D_\phi(s)$ \citep{ColesHarmon1989}.  With sufficient instantaneous S/N, the correlated amplitude 
and phase on individual baselines 
allow the structure function to be measured directly, providing access to 
the turbulence amplitude $C_N^2$, spectral index $\alpha$, and the 
inner scale $l_i$ through the form of General structure function (GSF) \cite{Ingale2015}  $D_\phi(s)$:
\begin{equation}
\label{eq:gsf}
D_\phi(s,\lambda) \;=\; 8\pi^2 r_e^2 \lambda^2 \int \Big[1-\tfrac{f_p^2(R)}{f^2}\Big] 
C_N^2(R)  \mathcal{K}_\alpha\!\Big(\tfrac{s}{l_i(R)}\Big)  dR,
\end{equation}
\revipz{In Eq.~\ref{eq:gsf}, $s$ is the interferometer baseline projected onto the scattering screen, $\lambda$ is the observing wavelength, $r_e$ is the classical electron radius, $f_p(R)$ is the plasma frequency at heliocentric distance $R$, $f$ is the observing frequency, $C_N^2(R)$ is the turbulence amplitude, $l_i(R)$ is the inner scale, and $\mathcal{K}_\alpha$ denotes the spectral form associated with the adopted power-law index $\alpha$.}
Longer baselines probe large-scale density fluctuations, while shorter baselines 
sample the turnover imposed by the inner scale; together, the multi-baseline visibility function constrains the full radial turbulence profile \cite{ColesHarmon1989}.

Multiplication of the visibility by $\Gamma(\mathbf{s})$ is equivalent, to a convolution of the intrinsic sky brightness 
$I_0(\boldsymbol{\theta})$ with a scattering kernel $G$, with Fourier transform:
\[
I(\boldsymbol{\theta})
   = I_0(\boldsymbol{\theta}) * G(\boldsymbol{\theta}),
\qquad
G = \mathcal{F}^{-1}\!\left[\Gamma(\mathbf{s})\right].
\]
Thus, the observed image encodes the same physical information as the 
visibility-domain coherence function.  
In the special and observationally common case where
$D_\phi(s) \propto s^2 $
the coherence function is Gaussian,
$\Gamma(s) = \exp(-a s^2)$
and hence the scattering kernel $G(\theta)$ is \emph{exactly Gaussian}.  
In this regime, the broadened source size (FWHM or second moment), axial ratio, and position angle provide a complete description of the scattering, and directly trace the turbulence amplitude and its anisotropy relative to the coronal magnetic field.

\subsection{The anisotropic of turbulence and magnetic field orientation.}
\revipz{In a magnetized corona, density fluctuations are typically anisotropic, with larger spatial correlation lengths along the background magnetic field than across it. As a result, radio-wave scattering is generally stronger across the field than along it, producing an anisotropic scatter-broadened image. This anisotropy is captured by imaging and can be reproduced by 3D stochastic/ray-tracing simulations \citep{kontar2023anis}.}

\revipz{The observed axial ratio, $\theta_{\rm maj}/\theta_{\rm min}$, quantifies the anisotropy of the scattering kernel, while the position angle traces the sky-projected magnetic-field orientation, subject to a 180$^\circ$ ambiguity.} \revipz{Accordingly, the image major axis is expected to be perpendicular to the local magnetic field, allowing measurements of $(e,\mathrm{PA})$ to trace $\mathbf{B}$ and its radial evolution.} Simulation-constrained values $k_{\parallel}/k_{\perp}\!\sim\!0.25$--$0.4$ reproduce observed sizes, time profiles, and apparent motions of near-Sun sources, and provide a forward-modeling link from measured $(e,\mathrm{PA})$ to the anisotropic turbulence spectrum and the heliospheric magnetic-field geometry \citep{kontar2023anis}.

\subsection{Meso- to large-scale density structures: higher moments and centroid shifts}

Meso- to large-scale coronal structures---such as streamers, coronal-hole boundaries, and other refractive gradients that act on scales much larger than those that produce scattering broadening. These low-$k$ fluctuations manifest in images as centroid shifts, asymmetric brightness profiles, and non-Gaussian higher moments. 
In the visibility domain they appear as smooth, baseline-dependent departures from a single-power-law structure function $D_\phi(s)$. Because refractive effects evolve slowly and scale weakly with frequency, measurements of centroid shifts, skewness, and kurtosis provide
a complementary diagnostic of mesoscale coronal structure. Together with visibility-derived turbulence parameters, these moments link small-scale scattering to broader density inhomogeneity along the line of sight.

\subsection{Moment-ordered source measurements}


As SKA's primary data product will be images, and the broadened radio source can be resolved in unprecedented detail. 
To make the standard observation products interpretable, we propose organizing them in the logical order of image moments, from the most basic (flux \& position) to progressively higher shape descriptors.

\paragraph{Source moments order:}
Let $I(x,y)$ be the sky intensity on a pixel grid (or continuous field).
Define raw moments $M_{pq}=\sum_{x,y} x^{p} y^{q} I(x,y)$ and the total flux
$M_{00}=\sum_{x,y} I(x,y)$. Centralize with the centroid
$\bar x = M_{10}/M_{00}$, $\bar y = M_{01}/M_{00}$ and let
$\mu_{pq}=\sum_{x,y} (x-\bar x)^p (y-\bar y)^q I(x,y)$ denotes central moments.

\begin{figure}[h]
  \centering
  \includegraphics[width=0.99\linewidth]{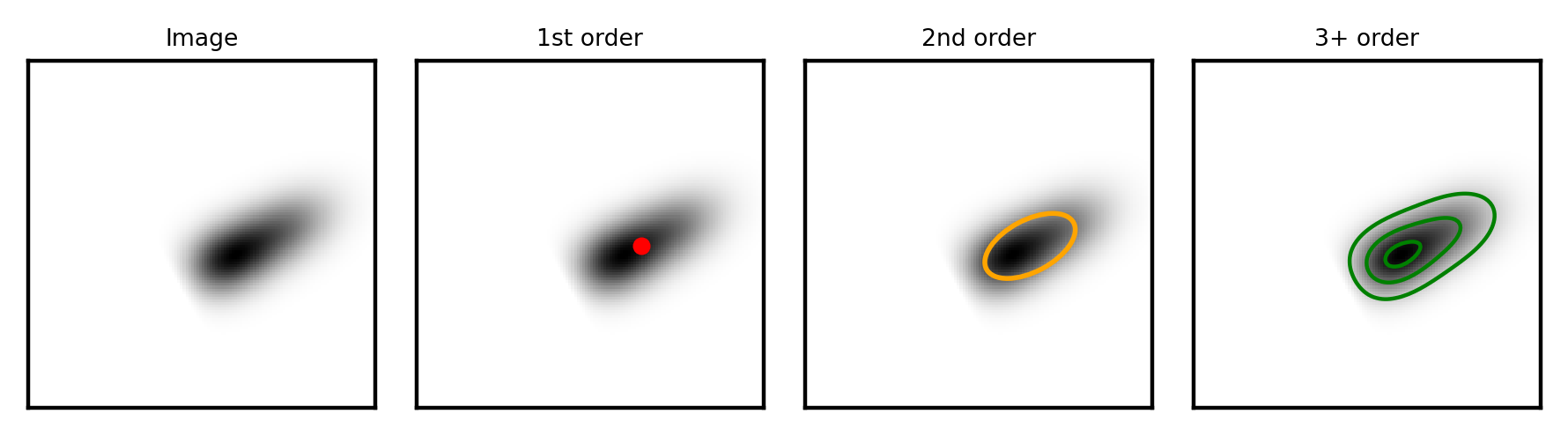}
  \caption{Concept demonstration of the higher order moments of the scattering kernel: 1st order (position offset), 2nd order (elongation), 3+ order (skew, kurtosis, and so on)}
  \label{fig:moment_concept}
\end{figure}

\begin{enumerate}
  \item[\textbf{1:}] \textbf{Position and total flux:}\
  \[
    M_{00} = \sum_{x,y} I(x,y), \qquad
    (\bar x,\bar y)=\Big(\frac{M_{10}}{M_{00}}, \frac{M_{01}}{M_{00}}\Big).
  \]
  \item[\textbf{2:}] \textbf{Size (FWHM) via a 2D Gaussian model:}\
  Use the $2\times 2$ covariance from the second central moments
  \[
    \Sigma=\frac{1}{M_{00}}
    \begin{bmatrix}
      \mu_{20} & \mu_{11}\\
      \mu_{11} & \mu_{02}
    \end{bmatrix}.
  \]
  Diagonalize $\Sigma$ to obtain $\sigma_{\rm maj},\sigma_{\rm min}$ and
  the orientation $\theta$. Report Gaussian FWHM along the principal axes:
  \[
    {\rm FWHM}_{\rm maj,min}=2\sqrt{2\ln 2}\;\sigma_{\rm maj,min}.
  \]
  \item[\textbf{3+:}] \textbf{Asymmetry/peakedness (skewness and kurtosis):}\
  Project onto the major axis coordinate and compute
  1D central moments $\mu_k=\sum I(t-\bar t)^k/M_{00}$:
  \[
    \text{Skewness}\ \gamma_1=\frac{\mu_3}{\mu_2^{3/2}}, \qquad
    \text{Kurtosis (excess)}\ \gamma_2=\frac{\mu_4}{\mu_2^{2}}-3.
  \]
  \revipz{Here $t$ denotes the coordinate obtained by projecting the image onto the major-axis direction of the fitted 2D Gaussian.}
\end{enumerate}

\begin{table}[h]
\centering
\caption{Moment-ordered measurements.}
\begin{tabular}{lll}
\hline
\textbf{Level} & \textbf{Definition (equation)} & \textbf{Description} \\
\hline
1 &
$M_{00}=\sum I,\quad (\bar x,\bar y)=\big(M_{10}/M_{00}, M_{01}/M_{00}\big)$ &
Total flux and centroid (position) \\
2 &
$\Sigma=\frac{1}{M_{00}}\begin{bmatrix}\mu_{20}&\mu_{11}\\\mu_{11}&\mu_{02}\end{bmatrix},\
{\rm FWHM}=2\sqrt{2\ln2} \sigma$ &
Source size/shape/PA as 2D Gaussian\\
3+ &
$\gamma_1=\mu_3/\mu_2^{3/2},\quad \gamma_2=\mu_4/\mu_2^{2}-3$ &
Asymmetry and peakedness \\
\hline
\end{tabular}
\end{table}

With this convention, by continuously taking the angular broadening observation, we can accumulate a catalog of multi-moment measurements. All previous works are restricted in the first and second orders. With the resolving power of SKA, it is possible to explore in detail the third and higher order moments.

\section{Statistics of sources near solar track}
\label{sec:2}


The previous angular-broadening work has relied on Tau A because it is exceptionally bright, compact, and located at low ecliptic latitude. To turn single-point experiments into a year-round probe of coronal/solar-wind turbulence, it is necessary to exploit the much larger population of {fainter} background sources that lie within a few to ten degrees of the solar track on any given day. SKA-Low's sensitivity and resolution make this practical: Jansky-scale sources become routine targets even in short integrations, enabling multi-source, multi-epoch mapping of scattering strength and anisotropy along the ecliptic. In this section we quantify the expected background sources along the Sun's path and summaries its brightness statistics, which set the observational requirements for SKA-Low angular broadening observations.

\begin{figure}[h]
    \centering
    \includegraphics[width=0.99\linewidth]{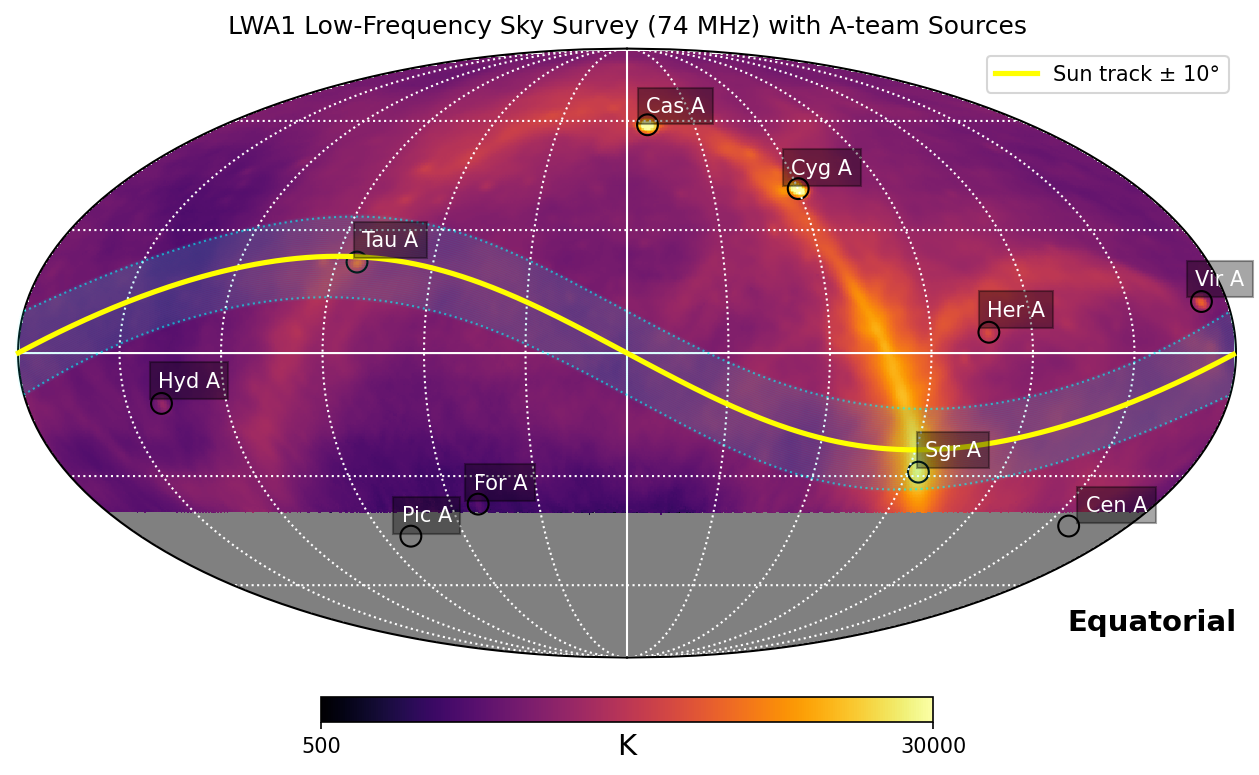}
    \caption{The Sun's annual track (yellow dashed line) on the all-sky map at 70 MHz derived from the LWA1 Low-Frequency Sky Survey \cite{dowell2017lwa1}. Prominent calibration (``A-team'') radio sources---including Cas A, Cyg A, Tau A, Her A, Vir A, Sgr A, For A, Pic A and Hyd A---are labelled for reference. The color scale shows the sky brightness temperature in kelvin.}
    \label{fig:skymap}
\end{figure}

\subsection{Sources near the solar track}

The Sun's annual path samples a wide range of sky environments (Fig. \ref{fig:skymap}), crossing the Galactic plane twice. Notably, \textit{Tau A} lies on the track; its annual conjunction provides an optimal window for angular-broadening studies because the source is extremely bright and compact---so even short integrations can resolve the anisotropically broadened image.

\begin{figure}
    \centering
    \includegraphics[width=0.85\linewidth]{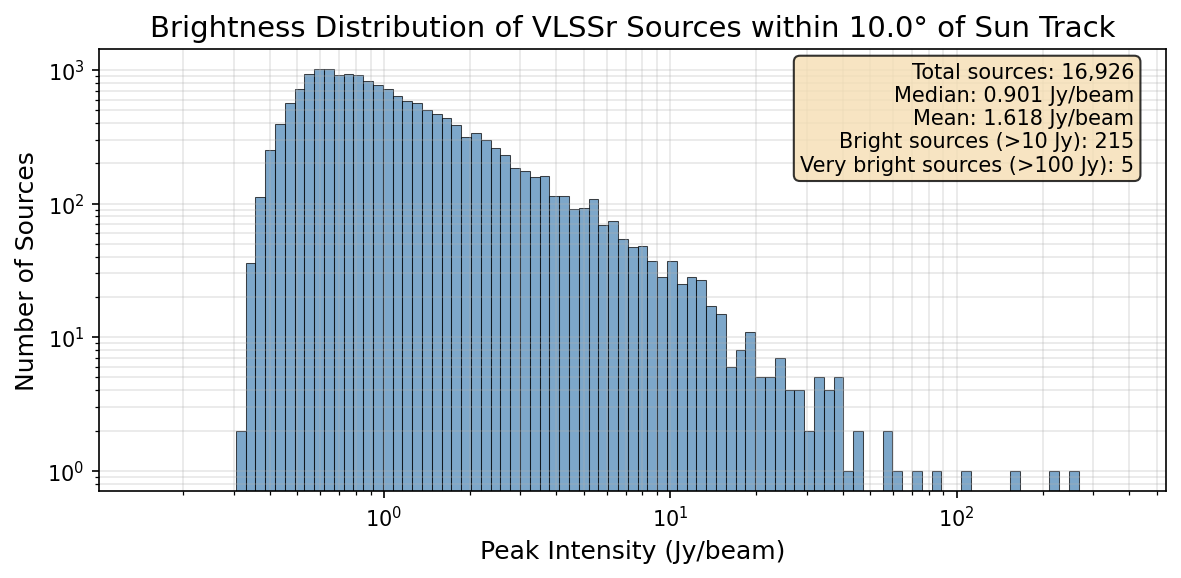}\\
    \includegraphics[trim=-1.7cm 0 0 0, width=1.015\linewidth]{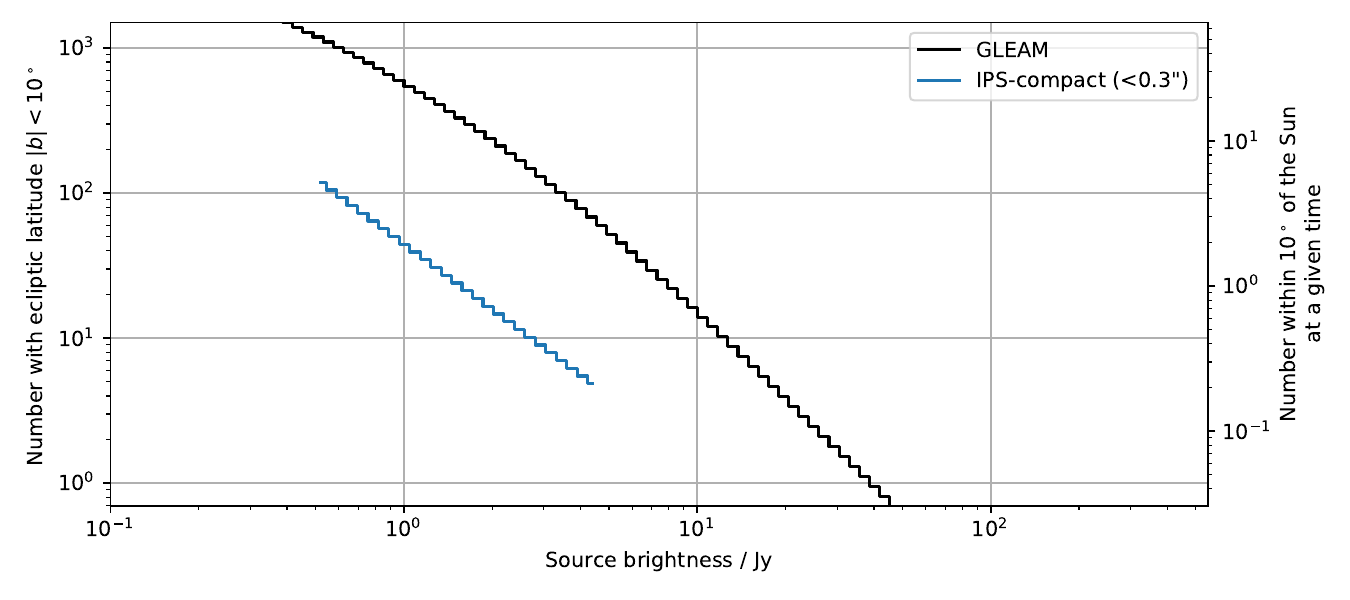}
    \caption{Top panel: Brightness distribution of VLSSr (74 MHz) sources located within 10$^\circ$ of the Sun's annual track. A total of 16,926 sources are included. About 215 sources exceed 10 Jy, and only 5 exceed 100 Jy. Bottom panel: Expected number of sources at each given flux density based on source counts of \citet{Franzen2019} based on low-frequency surveys \citep{Williams2016,Hurleywalker2017} and IPS source counts \citep{Chhetri2018}.}
    \label{fig:vlssr_hist}
\end{figure}

To assess how well SKA-Low can pursue angular broadening beyond this single beacon, we analyzed candidate background sources in the vicinity of the solar track, selecting objects within $\pm10^\circ$ (shaded band in Fig. \ref{fig:skymap}). A cross-match with the VLSSr catalogue at \SI{74}{MHz} (Fig. \ref{fig:vlssr_hist} upper panel) shows that very bright objects are rare along the ecliptic: only a few exceed \SI{100}{Jy}, whereas the population is dominated by sources at the $\sim$Jy level. This implies that deep, high-dynamic-range imaging of the near-Sun region is necessary to exploit the much larger reservoir of fainter calibrators/targets.

The turnover in source numbers evident in the upper panel of Fig. \ref{fig:vlssr_hist} below 0.5 mJy is due to the sensitivity of the survey. To predict numbers of sources below this limit, we use the source count statistics of \citet{Franzen2019} and \citet{Chhetri2018} to calculate the expected number of sources. These numbers agree well with VlSSr in the range of overlap, although the source count numbers are systematically low. This is due to the average low-frequency radio source having a spectral index\footnote{Most extragalactic sources have an inverse spectral energy distribution, such that their flux density $S$ at frequency $\nu$ is given by $S=S_0\left(\nu/\nu_0\right)^\alpha$, where $\alpha$ is the spectral index. Sources will be 75\% brighter at 75 MHz than at 150 MHz for $\alpha=-0.8$} of -0.8.

\revipz{Based on the source-count distribution in Fig.~\ref{fig:vlssr_hist}, a flux-density threshold of about $\sim$2 Jy yields of order tens of candidate sources within 10$^\circ$ of the Sun at a given time.} \citet{Kansabanik2022} have demonstrated sufficient dynamic range with the MWA to detect a 4.6 Jy source at $>$7-$\sigma$, however this requires exquisitely precise calibration. Nonetheless, this would allow tens of sources to be used as scatter broadening probes simultaneously, and further modest improvements could increase the number of usable sources by a further order of magnitude.

Another important consideration is source structure. Ideally, point-like sources would be used as any deviation from a point source can then be attributed to scattering effects.
For this reason, we have also included statistical source counts derived from IPS measurements \citep{Chhetri2018}.
\revipz{At the relevant frequencies, interplanetary scintillation is strongly suppressed for sources larger than about 0.3 arcsec, so IPS-selected sources can be treated as effectively point-like for SKA-Low angular-broadening studies} \citep{Chhetri2018}.
Roughly speaking, these make up one tenth of the low-frequency source population, so with high-quality calibration, a number of these sources will be usable at any given time.

\subsection{Temporal variation of the source statistics}

\begin{figure}
    \centering
    \includegraphics[width=0.90\linewidth]{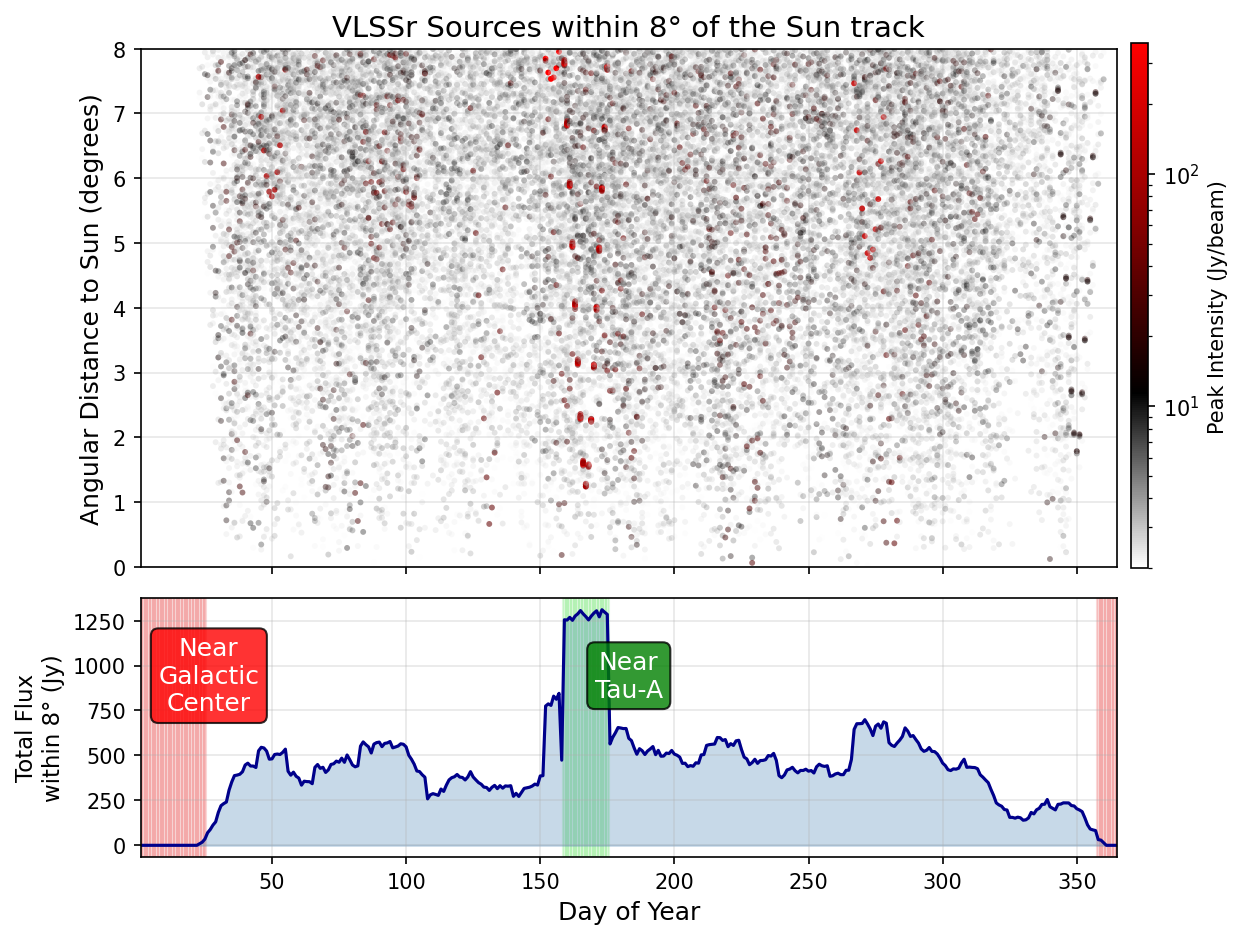}
    \caption{Distribution and integrated flux of VLSSr (74 MHz) sources located within 8$^\circ$ of the Sun's annual track \citep{2007AJvlaVLSS74}. \textbf{Top}: Each point represents a cataloged VLSSr source, color-coded by its peak intensity. \textbf{Bottom}: The total flux density summed over all sources within 8 Degree of the Sun as a function of day of year. The shaded regions highlight notable segments---an empty region close to the Galactic center not covered by VLSSr (red) and a high-flux region near Tau A (green)---illustrating how background source density and brightness vary along the solar path.}
    \label{fig:vlssr_doy}
\end{figure}

The analysis in the previous section considers extra-galactic sources and their average density across the whole sky. However, year-round temporal statistics further highlight the importance of \emph{when} to observe (Fig. \ref{fig:vlssr_doy}). During the first month of the year (DoY $\sim 0$--25), \revipz{where DoY denotes day of year,} the {total} flux within $10^\circ$ of the track at any given time can drop below 20 Jy, making this a poor period for angular-broadening measurements even with SKA-Low. By contrast, around DoY $\sim$155-175 the track passes near \textit{Tau A}, producing a strong peak in the summed background shown in \ref{fig:vlssr_doy}, and this period offers ideal conditions for angular-broadening experiments. For much of the remaining year the integrated flux within $10^\circ$ is typically of order a few hundred Jy ($\sim$500 Jy), providing workable, though less extreme, conditions. Taken together, the histogram and day-of-year curves provide a practical way to \emph{prioritize} observing windows for SKA-Low: schedule angular-broadening campaigns near the Tau A season and other high-flux intervals; reserve the lowest-flux segments for quiet-Sun studies and very deep integrations aimed at sub-Jy background sources.

\section{Sensitivity and resolution}
\label{sec:sensitivity}\label{sec:3}

\subsection{Sensitivity}

The thermal sensitivity of an aperture array depends on both its effective area and the (LST-dependent) system temperature,
\(A_{\rm e}(\nu)\) and \(T_{\rm sys}(\nu,{\rm LST}) = T_{\rm ant}(\nu,{\rm LST}) + T_{\rm rcv}(\nu)\) \citep[][Sec. 2.1]{2021SKA_AAVS_PASAsokoloski}. The antenna temperature is obtained by
beam-weighting the all-sky brightness \(T_{\rm sky}\) over the station beam, hence large changes with sky position (e.g. Galactic plane up/down) directly modulate sensitivity \citep[][Eqs. 1\&2]{2021SKA_AAVS_PASAsokoloski}; thus for the Galactic Centre region indicated in Fig. \ref{fig:vlssr_doy}, the sensitivity will be much lower, especially at lower frequencies. For SKA-Low the System Equivalent Flux Density (SEFD) follows the standard relation
\begin{equation}
{\rm SEFD} = \frac{2k T_{\rm sys}}{A_{\rm e}},
\end{equation}
with the Stokes\;I constructed from the X/Y giving
\({\rm SEFD}_I = \tfrac{1}{2}\sqrt{{\rm SEFD}_{XX}^2+{\rm SEFD}_{YY}^2}\) \citep[][Eq. (5)]{2021SKA_AAVS_PASAsokoloski}.
For a homogeneous array of \(N_{\rm st}\) stations, the array SEFD scales as the station SEFD divided by \(N_{\rm st}\) (for SKA1-Low \(N_{\rm st}=512\)):
\begin{equation}
{\rm SEFD}_{\rm array} = \frac{{\rm SEFD}_{\rm station}}{N_{\rm st}},
\label{eq:sefd_array}
\end{equation}
and the array sensitivity is often quoted as \(A_{\rm eff}/T_{\rm sys}\) using this array SEFD \citep[][Eqs. (3)--(5)]{macario2022characterization}. Commissioning measurements with the AAVS2 prototype show that SKA-Low sensitivity peaks when the Galactic plane is below the horizon and varies strongly with LST; measured values across 55--320 MHz agree with EM simulations and meet or exceed SKA-Low requirements by up to a factor of \(\sim2\) in favourable LST ranges \citep[][Sec. 4.2; Fig. 8]{macario2022characterization}.
For natural weighting and dual polarization, the interferometric radiometer equation gives the image thermal noise
\begin{equation}
\sigma_{\rm th} \simeq
\frac{{\rm SEFD}_{\rm array}}{\sqrt{n_{\rm pol} \Delta \nu  t}},
\quad (n_{\rm pol}=2).
\label{eq:radiometer}
\end{equation}

\subsubsection{Sensitivity away from the Sun}

AAVS2 measurements and simulations imply array sensitivities of order
\(A_{\rm eff}/T_{\rm sys}\sim 10^{3} {\rm m^2 K^{-1}}\) in the cold-sky LST ranges at 60--160 MHz (with clear LST dependence and modest X/Y differences) \citep[][Sec. 4.2 \& Fig. 8]{macario2022characterization,2021SKA_AAVS_PASAsokoloski}. As an example, at \(\nu=60\) MHz with a bandwidth of \(\Delta\nu=5\) MHz and an integration \(t=1\) s:
\begin{align}
A_{\rm eff}/T_{\rm sys} &= 1000 {\rm m^2 K^{-1}} \Rightarrow
{\rm SEFD}_{\rm array}\approx 2.76 {\rm Jy}.
\end{align}
Equation \ref{eq:radiometer} then gives
\[
\sigma_{\rm th}(1{\rm s},5{\rm MHz}) \approx
\frac{2.76\rm[Jy]}{\sqrt{2\times5{\times}10^{6} \rm{[Hz]} \times1[s]}} = 0.87 {\rm mJy}.
\]
Representing the weakest detectable brightness level, 
Scaling with time is \(\propto t^{-1/2}\): in 10 s one expects \(\sim0.3\text{--}0.5\) mJy, and in 60 s \(\sim0.1\text{--}0.2\) mJy.

\subsubsection{Sensitivity in the near vicinity of the Sun}

When the SKA-Low's zenith points toward \emph{hot} regions---with the Sun or near Galactic---the
system temperature rises dramatically. Measurements and simulations for aperture arrays show that
$T_{\rm sys}$ can increase by about {10 dB} relative to cold-sky LSTs (i.e.\ a factor of $\sim$10 in noise power;
see Fig. 18 of \citealt{benthem2021aperture}). Because ${\rm SEFD}\propto T_{\rm sys}$, the image thermal noise scales
linearly with this penalty. For our 60 MHz example with $\Delta\nu=5$ MHz and $t=1$ s
(cold-sky $\sigma_{\rm th}\!\approx\!0.9$ mJy beam$^{-1}$), a $10$-fold $T_{\rm sys}$ increase gives a
\textbf{near-Sun noise floor of $\sim$9 mJy beam$^{-1}$}. The quiet Sun typically has a brightness temperature
of $\sim\!10^{6}$ K and a flux density of $\sim\!10^{4}$ Jy at these frequencies
\citep{zhang2022imaging}; the \emph{theoretical} dynamic range (Sun peak divided by thermal noise) is therefore
\[
{\rm DR}_{\rm th}\;\sim\;\frac{10^{4}\;{\rm Jy}}{9\times10^{-3}\;{\rm Jy}}\;\approx\;1.1\times10^{6}.
\]
This vastly exceeds what calibration and imaging can usually deliver in practice, even with modern
direction-dependent calibration. As an illustration, Fig. \ref{fig:peelingdaytime} (processed with
DDC and solar/bright-source subtraction using \textsc{DP3}\footnote{https://dp3.readthedocs.io/en/latest/})
achieves an image dynamic range of $\sim\!630$, within which {multiple} $\sim$10 Jy-level
sources remain detectable. This gap between the thermal limit and achieved dynamic range underscores that near-Sun feasibility is determined primarily by {calibration accuracy and modeling of direction-dependent effects} rather than by raw sensitivity.
The image is produced with OVRO-LWA, 352 LWA element antenna. \revipz{Compared with OVRO-LWA, SKA-Low will provide substantially greater collecting area, baseline coverage, and calibration flexibility, which should significantly improve near-Sun imaging fidelity and the detectability of sub-Jy broadened sources.}

\subsection{Resolution}

\subsubsection{Angular resolution}
The ability to \emph{resolve} the scatter-broadened source shape is set by the synthesized beam
$\theta_{\rm b}\!\simeq\!\lambda/B_{\max}$. For SKA1-Low ($b_{\max}\!\approx\!65$ km, 50--350 MHz), this yields
$\theta_{\rm b}\!\approx\!16''$ at 60 MHz, and $\sim3''$ at 300 MHz.
With calibration and deconvolution weighting scheme, loss of resolution, we are expecting 2-5 times larger than theoretical value,  
which is still well below the expected near-Sun broadening (tens of arcsec to arcminutes at these frequencies \citealt{2025zhangTauA}), so the elongated source within $\sim$5--50 $R_\odot$ will be resolved and its axial ratios/position angles can be measured.
SKA-Mid ($b_{\max}\!\approx\!150$ km, $\gtrsim$350 MHz) delivers \revipz{synthesized beams}:
$\sim1.2''$ at 350 MHz, $\sim0.4''$ at 1 GHz, and $\sim0.14''$ at 3 GHz, enabling observations when
scattering is weaker and closer to the Sun (roughly 2--10 $R_\odot$) where $\theta_{\rm scat}\!\propto\!\nu^{-2}$
becomes small. Similar to SKA-Low, the working beam depends on the calibration and deconvolution weighting scheme;
even with robust weighting ($-0.5$--0), SKA-Low is expected to achieve $\sim$40'' at 60--80 MHz, while SKA-Mid remains
firmly sub-arcsecond above 1GHz. Thus the two arrays provide complementary resolution--frequency
coverage that together span the heliocentric ranges indicated in Fig. \ref{fig:concept}.


\subsection{Proof of concept with OVRO-LWA}

\begin{figure}[h]
  \centering
  \includegraphics[width=0.99\linewidth]{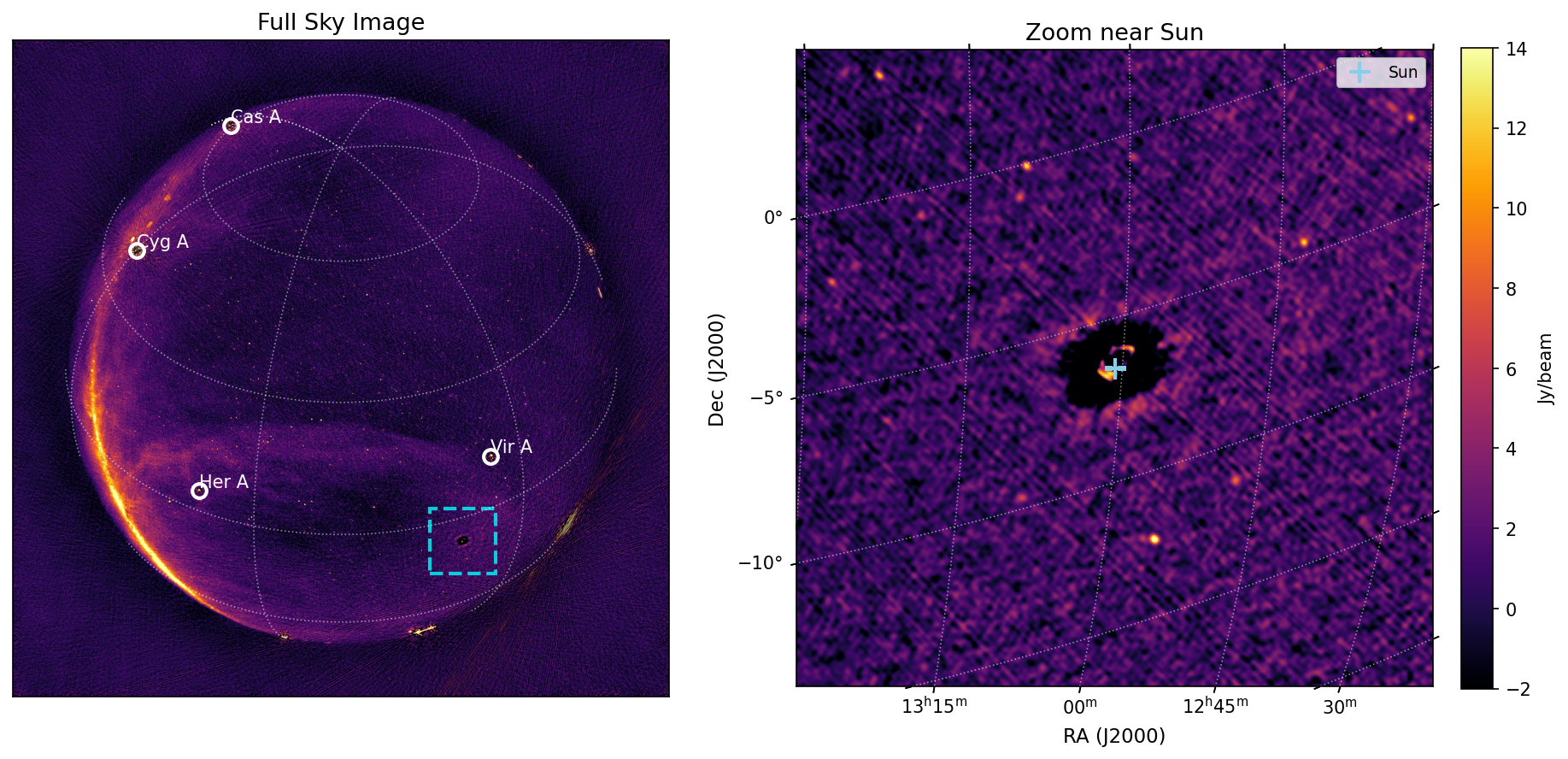}
  \caption{OVRO-LWA snapshot at 59 MHz ($\Delta\nu$ = 19.2 MHz) with a 10 s integration. Left: Full-sky image in equatorial coordinates; the Galactic plane is visible and bright ``A-team'' calibrators (Cas A, Cyg A, Her A, Vir A) are labeled. The cyan dashed box marks the region shown at right. Right: Zoom around the Sun (The Sun is modeled and subtracted); the cyan ``+'' indicates the solar ephemeris. The color scale is in Jy/beam. The compact solar emission and residual sidelobe structure illustrate near-Sun imaging performance in a short snapshot.}
  \label{fig:peelingdaytime}
\end{figure}

For example, at 60 MHz with $\Delta\nu=5$ MHz and $t=60$ s, cold-sky imaging reaches
$\sim\!0.1$--$0.2$ mJy beam$^{-1}$; even with near-Sun inflation factors of 10--20, residuals $\ll0.1$ Jy
are achievable after peeling, enabling detection of broadened $\sim0.1$--$1$ Jy targets at $5\sigma$ depending on scattering size (Sec. \ref{sec:sensitivity}).

\subsection{Proof of concept with MeerKAT}

As the precursor to SKA-Mid, recent MeerKAT L-band observations---conducted with the phase center placed about 2.5$^\circ$ from the Sun---have demonstrated the array's capability to detect compact-source angular broadening. Unlike previous studies that were limited to a small number of bright targets, the MeerKAT observations, benefiting from high sensitivity and a large effective field of view, notably reveal multiple broadened compact sources during the observation. This yields a sufficiently sampled dataset for preliminary statistical assessment, offering valuable insights into coronal density fluctuations and large-scale turbulence properties.

These angular-broadened sources are primarily located within the primary beam ($\approx 0.6^{\circ}$ at 1 GHz), where they are less attenuated and less affected by the bright solar signal, allowing more reliable analysis. Their intrinsic structures can be deconvolved from the synthesized beam to quantify the broadening characteristics, which in turn provide diagnostics of the local coronal conditions and magnetic topology at different locations relative to the Sun.

Through recent observations, MeerKAT has demonstrated its unique capability to detect multiple compact sources affected by angular broadening, serving as a valuable precursor for future SKA studies. By comparing with the VLBI catalog, most compact sources within the MeerKAT primary beam are successfully detected, although the broadening properties of some weaker or more distant sources remain difficult to characterize. This is mainly because, although MeerKAT already provides excellent sensitivity and resolution, they are still not sufficient for resolving all sources, particularly those farther from the phase center and less bright.

\begin{figure}[h]
    \centering
    \includegraphics[width=0.33\linewidth]{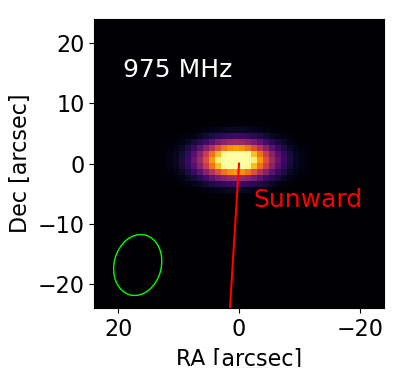}
    \caption{Left: MeerKAT observations of the the broadened source J1833-210, which has been deconvolved from the synthesized beam. Right: Simulations of the expected broadening of a point source in the presence of density fluctuations for \revipz{975 MHz} frequencies \citep{kontar2023anisotropic}. }
    \label{fig:MeerKAT_example}
\end{figure}

The most significant improvement expected from SKA will be its substantially higher angular resolution, enabled by much longer baselines, allowing it to resolve sources with smaller intrinsic and scattering sizes. In addition, improved \textit{uv} coverage and higher sensitivity will enhance the ability to image multiple compact sources simultaneously, even more close to the Sun. With these advances, SKA will be able to systematically observe a large population of compact sources over a wide range of heliocentric distances and position angles, providing comprehensive constraints on coronal scattering and turbulence. Such statistical coverage will represent a transformative step toward understanding the angular broadening phenomenon and the underlying plasma processes in the solar corona and solar wind.

The left panel of Fig. \ref{fig:MeerKAT_example} shows broadening of the point source J1833-210 as observed on 29 December 2024 above the solar pole with MeerKAT. The right panel shows simulations, using a modified version of the \citet{kontar2019anis} model, of how a point source, located $3R_{\odot}$ (solar radius, $R_{\odot}$) above the solar disk, will be broadened due to electron density fluctuations at a frequency of 1GHz.

\section{Density fluctuation cross-matching of in-situ and remote sensing}

\revipz{Inside the corona, remote-sensing techniques have constrained density fluctuations using radio-wave scattering, burst decay times, and direct radio imaging from multiple instruments and groups, including LOFAR and MWA precursor studies that estimate $\delta N_e/N_e$ from imaging \citep[e.g.][]{kontar2019anis,kontar2023anis,krupar2020density,murphy2021lofar,Mugundhan2017type3dens,McCauley2018,Mohan2021}.}
In the inner heliosphere, by contrast,
spacecraft provide direct measurements of $n_e$ and its fluctuations via QTN spectroscopy. Both avenues are powerful, but they have never sampled the \textbf{same plasma
volume at the same time}. With Parker Solar Probe (PSP) now advancing below $10 R_\odot$, and SKA-Low
delivering high-fidelity scattering measurements, we can, for the first time, cross-check
$\delta N_e/N_e$ from {remote sensing} against the {in situ} density spectrum in a common volume.


\begin{figure}[h]
    \centering
    \includegraphics[width=0.95\linewidth]{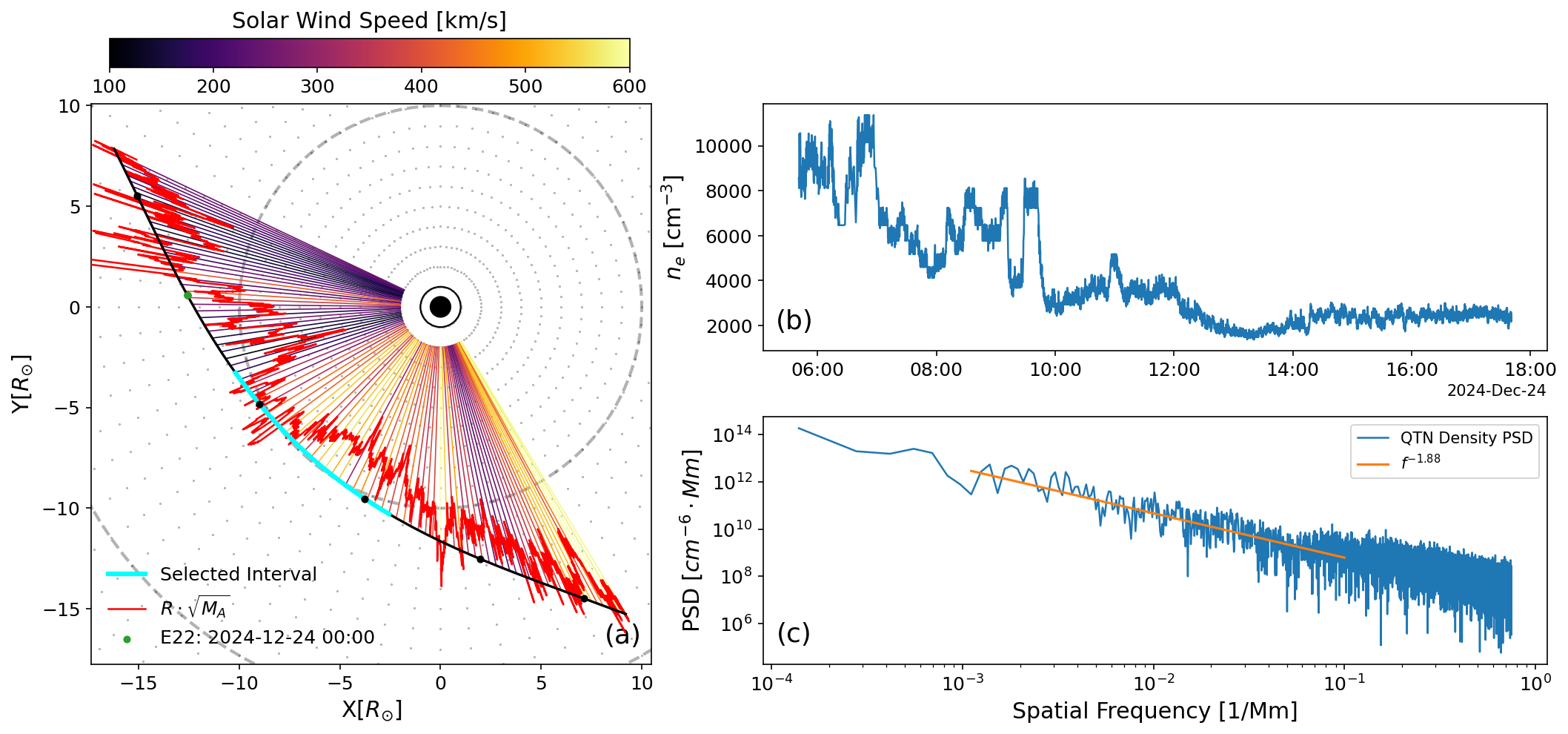}
  \caption{{Demonstration of in-situ measurement for perihelion \#22 (2024--12--24).}
  \emph{(a)} PSP trajectory in the near-Sun plane, colored by solar-wind speed. The cyan arc marks the
  tangent segment used to sample the background density. \emph{(b)} Quasi-thermal noise (QTN)-derived electron density $n_e(t)$ around
  perihelion. \emph{(c)} Power spectral density (PSD) of density fluctuations versus spatial frequency,
  with a power-law fit (slope $\approx -1.88$ shown in orange). These \textit{in situ} quantities can provide the $\delta N_e/N_e$ and spectral index against which the SKA-Low angular broadening inferences are
  validated.}
  \label{fig:psp_perihelion_demo}
\end{figure}

During PSP's perihelion passage (Fig. \ref{fig:psp_perihelion_demo}), we obtain
the in-situ electron density $n_e(t)$ and its density-fluctuation power spectral
density (PSD). The density is derived using the Quasi-Thermal-Noise method
\citep{moncuquet_first_2020}. As shown in panel (a), PSP traverses the
perihelion at $\sim9.86 R_\odot$ within a 12-hour interval during which the
solar wind remains sub-Alfv\'{e}nic ($M_A < 1$), placing the spacecraft inside the
Alfv\'{e}n surface \citep{kasper_parker_2021}. Panel (b) shows the temporal evolution
of $n_e$, and panel (c) gives the PSD of $n_e$ along the trajectory; the
spacecraft's nearly constant tangential speed allows spatial frequencies to be
interpreted at $\sim10 R_\odot$.

Simultaneously, SKA-Low provides angular-broadening measurements of background
sources. After calibration and peeling, the observed scattering kernel size,
axial ratio, position angle, and frequency scaling is fitted using a forward
model of wave propagation through a magnetized, turbulent corona. Multi-frequency
SKA-Low data constrain the fluctuation amplitude, while PSP supplies the absolute
density and local PSD. The comparison is made by matching the SKA-Low-derived
$\delta N_e/N_e$ at the tangent point with the PSP-based $\delta N_e/N_e$ and PSD
slope at the corresponding sector near $10 R_\odot$.

\textbf{Radio occultation (RO)} offers a complementary probe of coronal density fluctuations: spacecraft radio links grazing the Sun acquire frequency/phase scintillations that encode the line-of-sight integral of turbulence. Early \textit{Mariner/Helios/Viking} experiments established the technique \citep{Woo1976, Bird1982, Patzold1996}. \revipz{Recent radio-occultation studies using Akatsuki and the Mars Orbiter Mission have extracted quantitative turbulence spectra in the $\sim 4$--$10 R_\odot$ range. These studies found approximately Kolmogorov-like behaviour, radial variation in the spectral slope, and enhancements associated with coronal holes and streamers. Using frequency-residual spectra and Doppler spectral broadening of open-loop X/S-band links, they also derived solar-wind speed, electron density, and $\delta N_e/N_e$ \citep{Aggarwal2025a, Aggarwal2025b}. These results provide an important line-of-sight-integrated complement to SKA-Low scattering measurements.} Multi-epoch results mapped slow/fast wind between 1.4-10$R_\odot$; derived densities matched classic models \citep{Edenhofer1977, Strachan1993, Cranmer1999} but were reduced in studies like \cite{Doschek1997, Gallagher1999}.


RO thus complements the $\delta N_e/N_e$ measurements with LOS-integrated constraints for the density and the column density fluctuation. Joint analysis of SKA-Low scattering measurements, PSP in-situ, and RO measurements will enable direct cross-checks of $\delta N_e/N_e$, turbulence spectra, and radial evolution. Providing an anchor to the model for how coronal turbulence feeds the solar wind.

\section{Observation plans and strategies}
\label{sec:4}

The observation strategy is source and activity dependent, tailors time resolution, and calibration depth to the \emph{instantaneous background source densities} within $\pm 10^\circ$ of the solar track (Sec. \ref{sec:2}). 
We use the summed catalogued flux (e.g.\ VLSSr) as a proxy for the likelihood of finding multiple compact targets in the field.
Below we outline four regimes and the corresponding observing modes.

\textbf{Common practices across regimes}
\begin{itemize}
  \item Direction-dependent calibration (DDC) + bright source peeling every 1--5 s during active periods (Sec. \ref{sec:2},  Fig \ref{fig:peelingdaytime}).
  \item {Scheduling by DoY:} use the year-round flux curves (Fig. \ref{fig:vlssr_doy}) to prioritize Tau A season and other high-yield windows.
  \item {Data Products:} (i) high-time-resolution image cubes for Tau A season; (ii) deep, long-baseline images (and visibility-domain fits) of broadened targets; (iii) per-epoch catalogs of size/axial ratio/PA vs.\ elongation.
\end{itemize}

For the source-dependent strategy, we categorize the observation strategy as shown below. 
\subsection*{(A) Tau A solar conjunction: high-cadence campaign}
When near the annual Tau A conjunction (DoY $\sim$155--175). 
{Rationale is that} Tau A provides an ultra-bright, effectively point-like beacon that traverses a range of solar elongations and position angles within days.\\
\textbf{Observation and processing strategy:} (1) Time resolution $\Delta t \!\lesssim\!0.1$ s.
(2) Weighting uniformly to get higher spatial resolution.\\
\textbf{Science focus:} resolve \emph{temporal} variations of the scattering kernel $($size, axial ratio, PA$)$ on sub-second scales, constraining turbulence dynamics and anisotropy close to the Sun.

\subsection*{(B) Preferred background: $\sum S_{10^\circ}\gtrsim 400$ Jy}
Near ecliptic segments with many bright sources (outside Tau A peak).\\
\textbf{Observation and processing strategy:} 
Time resolution $\Delta t = 1$--5 s. (2) Weighting Briggs $-0.5$ to give relative higher weight to shorter baselines to achieve higher dynamic range.\\
\textbf{Science focus:} routine angular-broadening measurements on multiple Jy-level sources per integral; map anisotropy PA around the Sun.

\subsection*{(C) Moderate background: $\sum S_{10^\circ} \in (50\text{, \ }400 )$ Jy}
During most common part of the annual track.\\
\textbf{Observation and processing strategy:} 
(1) Deep imaging with $\Delta t=10$--30 s snapshots to push to sub-Jy targets; iterate DDC on 10--30 s intervals.
(2) Agreesive flagging; preserve raw channels until peeling is complete.
(3) Use weighting briggs 0-0.5 to emphasize more shorter baselines. \\
\textbf{Science focus:} increase source counts to build 2D maps of scattering strength and anisotropy, and to trace coronal magnetic-field orientation via the elongation PA.

\noindent This source-dependent plan ensures that we exploit the unique bright-beacon opportunities (Tau A),
harvest routine multi-source constraints in typical fields, and dedicate the cleanest windows to quiet-Sun
and calibration validation---maximizing science return over the full solar year.

\subsection{Visibility Data Availability}

\cite{Jaradat2025PASAvlbaJ} using 12-m class dishes at SKA-Mid frequencies, showed that a single VLBI baseline can not only detect scatter-broadening–induced decoherence, but also track the underlying phase fluctuations that produce the broadened ensemble image. 
Similar SKA observations would yield a rich dataset bridging current VLA and VLBI constraints, enabling the phase structure function to be mapped in 2D space and time. Crucially, this requires access to raw visibilities on each individual baseline, rather than only gridded visibilities, underscoring how powerful baseline-level data are as a probe of coronal scattering.

As shown in Sect.~\ref{sec:1} and demonstrated by \citep{Jaradat2025PASAvlbaJ}, visibilities are the natural domain for measuring scatter broadening and related higher-order effects. Images are one step further removed from the fundamental quantities that we are trying to observe. Therefore, at least in some cases we will need to request visibilities in order to get the full value from our observations. It may be possible to recover these by re-inversion of a dirty image, but experimentation would be necessary.

\section{Summary and outlook}
\label{sec:summary}

Angular broadening of compact background sources is a sensitive line-of-sight probe of coronal and solar-wind turbulence (fluctuation amplitude, anisotropy, dissipation scale, and magnetic field orientation). Using all-sky surveys and SKA performance estimates, we quantified year-round opportunities and the calibration/sensitivity required for routine measurements.
We have the following {key results}: 
\begin{itemize}
  \item \emph{Source availability along the solar track:} a cross-match with VLSSr shows that very bright objects are rare near the ecliptic, but thousands of $\sim$Jy sources lie within $\pm10^\circ$ of the Sun through the year, enabling multi-epoch, multi-source campaigns.
  \item \emph{Sensitivity and Resolution:} cold-sky array sensitivities of $A_{\rm eff}/T_{\rm sys}\!\sim\!10^3 {\rm m^2 K^{-1}}$ correspond to $\sigma_{\rm th}\!\approx\!0.9 {\rm mJy beam^{-1}}$ in a 1 s, 5 MHz snapshot at 60 MHz; hot-sky conditions inflate $T_{\rm sys}$ by $\sim$10 dB, yielding $\sim$9 mJy beam$^{-1}$. Practical limits are therefore set by dynamic range and calibration rather than raw sensitivity. SKA1-Low ($b_{\max}\!\sim\!65$ km) resolves tens-of-arcsec to arcmin scatter-broadened kernels across 50--350 MHz; SKA-Mid extends to \revipz{synthesized beams} at $\gtrsim$350 MHz, providing complementary leverage on smaller heliocentric distances.
  \item \emph{Calibration:} Direction-dependent calibration and solar/bright-source peeling are essential. With high-fidelity DDC, broadened $0.1$--$1$ Jy targets within a few degrees of the Sun are detectable on 10s timescales.
\end{itemize}

With the unbalanced background source distribution, we propose the observation strategy as {Source-dependent scheduling:} (i) high-cadence ($\Delta t\!\lesssim\!0.1$ s) campaigns near Tau A conjunction; (ii) 1--5 s imaging in high-background windows ($\sum S_{10^\circ}\!\gtrsim\!400$ Jy); (iii) 10--30 s deep snapshots in moderate backgrounds ($50$--$400$ Jy).
The resulting data products would be per-epoch catalogs of 1-3 orders of moments which can be used for further density fluctuation and turbulence studies.

High sensitivity provides long duration of availability of the angular broadening observation, with which we can perform {cross-validation between remote sensing and in-situ observations}, when the Earth-source line of sight is tangent to the PSP perihelion sector, SKA scattering measurements of $\delta N_e/N_e$ can be directly compared with \textit{in situ} density spectra from QTN, anchoring remote-sensing in the inner corona. LOS-integrated density and fluctuation constraints from spacecraft \revipz{RO will complement} SKA and PSP, enabling joint inversions of fluctuation amplitude, spectrum, and radial evolution.

With SKA1-Low/Mid, year-round, high-fidelity angular-broadening measurements \revipz{will become routine}. The programme will deliver: (i) a multi-year map of coronal anisotropy as a magnetic tracer, (ii) constraints on the radial evolution and dissipation of turbulence, and (iii) direct cross-checks of $\delta N_e/N_e$ between remote and \textit{in situ}/RO techniques. These outcomes will provide quantitative boundary conditions for heliospheric models and new insight into how coronal turbulence seeds and heats the solar wind.

\bibliographystyle{abbrvnat-maxbibnames4}
\bibliography{chapter} 

\appendix

\section*{Appendix}

\begin{longtable}{ccccc}
\caption{Bright (>15 Jy/beam) Sources within 5$^\circ$ of Solar Track from VLSSr}
\\
\hline
Rank & RA (hms) & Dec (dms) & Peak Intensity (Jy/beam) & DoY \\
\hline
1 & 05:34:33.28 & +22:02:01.4 & 256.648 & 162 \\
2 & 05:34:26.65 & +22:03:39.3 & 212.018 & 162 \\
3 & 12:29:05.89 & +02:02:55.7 & 161.687 & 272 \\
4 & 05:34:26.35 & +22:01:47.7 & 81.965 & 162 \\
5 & 05:34:41.32 & +22:01:34.8 & 75.072 & 163 \\
6 & 05:34:34.73 & +22:03:40.3 & 64.482 & 162 \\
7 & 05:34:21.39 & +22:03:59.5 & 43.981 & 162 \\
8 & 06:13:49.81 & +26:04:38.6 & 43.430 & 173 \\
9 & 08:53:08.57 & +13:52:54.3 & 40.210 & 214 \\
10 & 05:02:58.55 & +25:16:24.6 & 38.601 & 156 \\
11 & 03:10:00.01 & +17:05:57.4 & 38.428 & 127 \\
12 & 00:37:03.97 & -01:09:10.7 & 36.021 & 87 \\
13 & 12:56:11.16 & -05:47:23.8 & 35.515 & 278 \\
14 & 06:45:24.09 & +21:21:45.2 & 34.690 & 180 \\
15 & 09:42:15.30 & +13:45:49.3 & 32.476 & 224 \\
16 & 13:38:07.90 & -06:27:11.7 & 32.059 & 289 \\
17 & 10:10:59.99 & +06:24:45.5 & 31.802 & 237 \\
18 & 10:08:00.12 & +07:30:16.2 & 30.986 & 234 \\
19 & 09:44:15.61 & +09:46:13.5 & 28.537 & 228 \\
20 & 09:50:10.63 & +14:19:59.5 & 27.775 & 226 \\
21 & 00:06:22.61 & -00:04:26.5 & 26.516 & 77 \\
22 & 10:42:44.57 & +12:03:33.2 & 25.452 & 241 \\
23 & 21:23:01.72 & -16:28:02.6 & 25.143 & 34 \\
24 & 22:25:47.42 & -04:57:02.9 & 25.116 & 55 \\
25 & 12:42:19.60 & -04:46:20.5 & 24.925 & 274 \\
26 & 21:16:36.42 & -20:55:51.3 & 24.575 & 35 \\
27 & 04:56:43.32 & +22:49:23.3 & 24.138 & 153 \\
28 & 06:04:28.52 & +20:21:18.1 & 24.093 & 171 \\
29 & 09:06:31.81 & +16:46:14.4 & 22.853 & 215 \\
30 & 08:58:41.47 & +14:09:42.4 & 22.206 & 215 \\
31 & 10:26:32.19 & +06:27:28.0 & 21.443 & 239 \\
32 & 13:47:01.66 & -08:03:21.9 & 20.906 & 291 \\
33 & 06:43:07.36 & +23:19:02.3 & 20.154 & 179 \\
34 & 08:40:47.75 & +13:12:24.6 & 19.731 & 214 \\
35 & 17:30:38.23 & -21:28:45.0 & 19.179 & 346 \\
36 & 08:05:33.30 & +24:10:07.9 & 19.068 & 200 \\
37 & 01:18:18.44 & +02:58:06.6 & 18.811 & 99 \\
38 & 07:29:27.99 & +24:36:23.5 & 18.481 & 191 \\
39 & 23:28:12.22 & -04:56:11.9 & 18.208 & 67 \\
40 & 23:25:19.55 & -04:57:39.3 & 18.167 & 66 \\
41 & 12:32:00.07 & -02:24:02.6 & 18.064 & 271 \\
42 & 14:27:38.11 & -12:03:40.2 & 17.525 & 302 \\
43 & 11:02:17.45 & +10:29:07.1 & 17.339 & 246 \\
44 & 16:46:05.21 & -22:28:02.4 & 16.598 & 336 \\
45 & 12:06:19.85 & +04:06:10.9 & 16.549 & 265 \\
46 & 14:02:08.75 & -15:10:08.4 & 15.469 & 297 \\
47 & 11:41:08.28 & +01:14:19.3 & 15.355 & 257 \\
48 & 08:54:39.15 & +14:05:52.2 & 15.170 & 214 \\
49 & 08:35:03.84 & +14:11:49.4 & 15.070 & 210 \\
50 & 06:52:48.44 & +22:32:23.3 & 15.014 & 181 \\
\hline
\label{tab:top50_sources}
\end{longtable}

\end{document}